\documentclass[prd,showpacs,preprintnumbers,amsmath,amssymb,superscriptaddress,nofootinbib]{revtex4}
\usepackage{graphicx}
\usepackage{dcolumn}
\usepackage{bm}
\usepackage{graphicx,amsmath,amssymb,color}
\usepackage{relsize}
\usepackage{slashed}
\usepackage{graphicx}
\usepackage{bm}
\usepackage{mathtools}
\usepackage{mathrsfs}
\usepackage{soul}

\usepackage{verbatim} 
\usepackage{tikz}
\usepackage{array}
\usepackage{float}
\usepackage{hyperref}
\usepackage[caption=false]{subfig}

\newcommand{\half}[1][1] {\mathsmaller{\frac{#1}{2}}}

\allowdisplaybreaks
\def\be{\begin{eqnarray}}
\def\ee{\end{eqnarray}}
\def\bea{\begin{eqnarray}}
\def\eea{\end{eqnarray}}
\begin{document}
\title{Form Factors and Generalized Parton Distributions \\ in Basis Light-Front Quantization}
\author{Lekha Adhikari}\email{adhikari@iastate.edu}
\affiliation{Department of Physics and Astronomy, Iowa State University,
Ames, IA 50011, U.S.A.}
\author{Yang Li}\email{leeyoung@iastate.edu}
\affiliation{Department of Physics and Astronomy, Iowa State University,
Ames, IA 50011, U.S.A.}
\author{Xingbo Zhao}\email{xbzhao@impcas.ac.cn}
\affiliation{Institute of Modern Physics, Chinese Academy of Sciences, Lanzhou 730000, China.}
\author{Pieter Maris}\email{pmaris@iastate.edu}
\affiliation{Department of Physics and Astronomy, Iowa State University,
Ames, IA 50011, U.S.A.}
\author{James P. Vary}\email{jvary@iastate.edu}
\affiliation{Department of Physics and Astronomy, Iowa State University,
Ames, IA 50011, U.S.A.}
 \author{Alaa Abd El-Hady}\email{alaa\_abdelhady@yahoo.com}
\affiliation{Physics Department, Zagazig University,
 Zagazig 44519, Egypt.}
\date{\today}
\begin{abstract}                             
We calculate the elastic form factors  and the Generalized Parton Distributions (GPDs) for four low-lying bound states of a demonstration fermion-antifermion system, strong coupling positronium ($e \bar{e}$), using Basis Light-Front Quantization (BLFQ). Using this approach, we also calculate the impact-parameter dependent GPDs $q(x, {\vec b_\perp})$ to visualize the fermion density in the transverse plane (${\vec b_\perp}$). We compare selected results with corresponding quantities in the non-relativistic limit to reveal relativistic effects. Our results establish the foundation within BLFQ for investigating the form factors and the GPDs for hadronic systems.
\end{abstract}
\maketitle
\section{Introduction}
Form Factors (FFs) are among the most important measurable quantities that provide information on the internal structure of hadrons. Generalized Parton Distributions (GPDs) have been introduced as an additional tool to describe hadronic substructures. Unlike Parton Distribution Functions (PDFs), which depend only on the momentum fraction $x$, GPDs also depend on the momentum transfer $\Delta$. GPDs are defined as the non-forward matrix elements of the  same light-cone operators whose forward matrix elements (i.e. the expectation value) yield the PDFs \cite{mb:IJMP}. In specific kinematic regions, GPDs yield the conventional FFs \cite{Ji:PRL}. Therefore, GPDs are hybrid quantities having features in common both with the FFs and with the PDFs.  To gain a comprehensive understanding of physics  that underlies FFs, one needs the decomposition of FFs with respect to the momentum fraction of the active parton (quark) that absorbs the photon. GPDs, by definition, provide FF decompositions evaluated at a given value of the invariant momentum transfer $t = \Delta^2$. This feature, known as the GPD sum rules, allows one to calculate momentum-dissected FFs using the same  covariant current operator in Quantum Electrodynamics (QED) and Quantum Chromodynamics (QCD) \cite{mb:GPD, Diehl:2003ny}.

Over more than a decade, there has been a strong interest in GPDs as many observables can be linked with them. Specifically,
 GPDs have been used extensively to investigate the total angular momentum of the quarks/gluons within a hadron, thus forming the foundation for the field of spin physics \cite{Ji:PRL}. Moreover, they have been used to visualize hadrons in  three dimensions after performing suitable Fourier transforms \cite{mb:GPD, Diehl:2002he}. The resulting images are conveniently presented in a space where one dimension describes the light-cone
momentum fraction ($x$) and the other two dimensions describe the transverse position (${\vec b}_\perp$) of the parton (relative to the
transverse center of momentum). These distributions in the transverse plane also preserve the partonic interpretations. Further details of other hadronic correlation functions can be found in Refs. \cite{mb:GPD, Diehl:2003ny, Diehl:2002he}. Even though GPDs cannot be measured directly from experiments, they enter the  Deeply Virtual Compton Scattering (DVCS) amplitude through convolution integrals. The real and imaginary part of the DVCS amplitude can be separated in experiments using the beam charge and beam spin asymmetry, respectively \cite{Ji:1996nm, Radyushkin:1997ki, Brodsky:2000xy, Diehl:2000xz, Brodsky:2006in, Brodsky:2006ku}.

Several investigations \cite{Brodsky:2007hb, Frederico:2009fk, Maris:2000sk, deMelo:1997cb, deMelo:2002yq, Braguta:2007fj, Mezrag:2015xia} have presented  the FFs and  GPDs for quark-antiquark bound states. For example, in Ref.~\cite{Frederico:2009fk}, the pion GPDs and electromagnetic (e.m.) FFs have been calculated using two Light-Front (LF) phenomenological models (Mandelstam-inspired LF Model and LF Hamiltonian Dynamics model). Similarly, in Ref.~\cite{deMelo:2002yq}, the e.m.  FF is calculated for a two-fermion, pion-like system in the Breit frame.

We are motivated by these previous works to evaluate the elastic FFs of a demonstration fermion-antifermion system, strong coupling positronium ($e \bar{e}$), using the overlap integrals between light-front wavefunctions (LFWFs) in the Drell-Yan frame. Within the same frame, we also calculate the GPDs using the same LFWFs that were used to calculate the FFs. Positronium at strong coupling can be viewed as a prototype of quark-antiquark quarkonium systems, e.g., $c \bar c, b\bar b$. In the present work, we calculate the elastic FFs and GPDs for the leading Fock sector $|e\bar{e} \rangle$. Our current FF and GPD results serve as prototypes for future applications to quarkonium systems that are solved in the same (non-perturbative) bound-state  framework \cite{Li:2015zda}.

 In this work, we adopt Basis Light-Front Quantization (BLFQ), a recently developed $\textit{ab initio}$ approach  \cite{Vary:2009gt}, to calculate FFs and GPDs for four low-lying bound states of positronium $1{}^1\!S_0\, (0^{-+})$, $1{}^3\!S_1\, (1^{--})$, $2{}^1\!S_0\, (0^{-+})$, and $2{}^3\!P_0\, (0^{++})$. Here, states are identified with their non-relativistic quantum numbers (relativistic quantum numbers) $N{}^{2S+1}\!L_J\, (J^{PC})$, where $N$ is the principal quantum number\footnote{The relation between $N$, the principal quantum number, and $n$, the radial quantum number used in Particle Data Group, is $N=n+L$.}, $L$ is the total orbital angular momentum, $S$ is the total intrinsic spin, $J$ is the total angular momentum, $P$ is the parity and $C$ is the charge conjugation. BLFQ is a non-perturbative approach for solving bound state problems in quantum field theory \cite{Wiecki.2014,Vary:2009gt}. It is a Hamiltonian-based method \cite{Vary:2009gt} that combines the advantages of light-front dynamics \cite{Brodsky:1997de} with recent advances in nuclear many-body calculations \cite{Navratil:2000ww}. BLFQ has been successfully applied \cite{Honkanen:2010rc, Zhao:2014xaa} to the single electron problem in QED in order to evaluate the anomalous magnetic moment of the electron. Another recent work \cite{Chakrabarti:2014cwa} has presented the electron GPD calculations as a test problem for the BLFQ approach. Furthermore, the BLFQ approach has been extended to time-dependent strong external field problems such as non-linear Compton scattering \cite{ Zhao:2013cma, Zhao:2013jia}.  The FFs, GPDs, and impact-parameter dependent GPDs for the bound states of positronium at strong coupling in the BLFQ approach are the main results of this paper.

 We organize this paper as follows. In Sec. (II), FFs and GPDs are defined using the overlap integrals between LFWFs in relative coordinates. In Sec. (III), we briefly introduce the BLFQ approach, and in Sec. (IV), we present our results for FFs, GPDs, and impact-parameter dependent GPDs. Finally, we present the summary and outlook in Sec. (V).
\section{Form Factors and Generalized Parton Distributions on the Light Front}
\hspace{\parindent}
The elastic FFs are defined as {\cite{Drell:1969km, Brodsky:2007hb, Frederico:2009fk, Brodsky:1997de}}
\be I_{m_J,m_J'}(t \equiv \Delta^2) \triangleq  \frac{1}{2P^+}\langle \psi_{m'_J}^{J*}(P') |j^+(0)|\psi_{m_J}^J( P) \rangle, \ee
where  $P$ and $P'$ are initial and final state momenta of the system, respectively, $\Delta \equiv P'-P$ is the momentum transfer (we choose the Drell-Yan frame $ \Delta^+=0, t \equiv \Delta^2 = -{\vec \Delta}^2_\perp <0$),\, $j^{\mu}$ is the current operator, $J$ is total angular momentum of the system and  $m_J$ is the total angular momentum projection for the system. For simplicity, the charge of the electron $e$ is excluded in the definition of the FFs.

For $J = m_J=0 $, the above relation directly produces the charge FF ($G_C$). But for $J =1 $ in LF dynamics, due to the light-front parity and the charge conjugation symmetries, we may define four independent helicity amplitudes using the nine elastic FFs $I_{m_J,m_J'}$ with $m_J$ (and $m_J'$)$\,= 1, 0, -1$. For example, conventional FFs such as the charge FF ($G_C$), magnetic FF ($G_M$), and quadrupole FF ($G_Q$) can be computed using these amplitudes \cite{Brodsky:2006ez,Sachs:1962zzc, Yennie:1957, Grach:1983hd, Lorce:2009bs, deMelo:1997hh, Li:2015zda}. For simplicity, in the present study we limit the FF cases to $I_{0,0}(t)$ for $J=0,1$ which are sufficient to have a direct comparison between states with different $J$ \cite{Grach:1983hd,deMelo:1997hh}.

In the present work, we consider the limited case where the virtual photon couples only  to the  electron to calculate the helicity non-flip GPDs and corresponding FFs. Up to the leading Fock sector $|e \bar{e} \rangle$, within the impulse approximation, the elastic FFs using the Drell-Yan formula read
\be F(t) \triangleq I_{0,0}(t)= \sum_{\lambda_e,\lambda_{\bar e}}\int dx_e \int d^2 \vec k_\perp \, \psi^*(\vec k'_\perp, x_e, \lambda_e,\lambda_{\bar e}) \psi(\vec k_\perp, x_e, \lambda_e,\lambda_{\bar e}), \label{eq:form_factor_intrinsic}  \ee
where ${\vec k}_\perp$  and  $\vec{k'}_\perp=\vec{k}_\perp+(1 - x_e){\vec \Delta}_\perp$ are the respective relative transverse momenta of the electron before and after being struck by the virtual photon, $x_e$ ($x_{\bar{e}}$) is the longitudinal momentum fraction of the electron (positron) satisfying $x_e +x_{\bar{e}}=1$, ${\vec \Delta}_\perp$ is the transverse component of the momentum transfer, and $\lambda_e$ ($\lambda_{\bar e})$ is the spin of the electron (positron). The LFWF $\psi(\vec k_\perp, x_e, \lambda_e,\lambda_{\bar e})$ is normalized according to
\be \sum_{\lambda_e, \lambda_{\bar e}} \int dx_e \int d^2 {\vec k}_{\perp} \big | \psi(\vec k_\perp, x_e, \lambda_e,\lambda_{\bar e}) \big |^2= 1  \ee and for simplicity, we have suppressed the quantum numbers labeling $\psi$.

The helicity non-flip GPDs in the region $0\leq x_e \leq 1$ can be written as overlap integrals between LFWFs \cite{Frederico:2009fk, Brodsky:2000xy, Diehl:2003ny}
\be H(x_e,\xi=0, t=-{\vec \Delta}_\perp^2) =  \sum_{\lambda_e,\lambda_{\bar e}}
\int d^2 \vec k_\perp \, \psi^*(\vec k'_\perp, x_e, \lambda_e,\lambda_{\bar e}) \psi(\vec k_\perp, x_e, \lambda_e,\lambda_{\bar e}), \label{eq:gpds_intrinsic}
\ee
where $\xi \equiv - \Delta^+ / (P'^+ + P^+) $ is the skewness parameter and in $\Delta^+=0$, $\xi = 0$.

It is straightforward to extend our framework to helicity-flip GPDs, and (as mentioned before) to $q\bar{q}$ bound states, but for simplicity, we consider only four low-lying bound states of positronium here, for which there are well-established results in the non-relativistic limit.

In the Drell-Yan frame, the expressions for the GPDs (Eq.~\ref{eq:gpds_intrinsic}) are very similar to the expressions for FFs, except that the longitudinal momentum fraction $x_e$ of the electron is not integrated over. Therefore, GPDs defined  in Eq.~\ref{eq:gpds_intrinsic} are also known as ``momentum-dissected FFs" and  measure the contribution of the electron with momentum fraction $x_e$ to the corresponding FFs in Eq.~\ref{eq:form_factor_intrinsic}.

Now, referring to Ref.\cite{mb:GPD}, the impact-parameter dependent GPDs are defined as the Fourier transform of the GPDs with respect to the momentum transfer $\Delta$
\be  q(x \equiv x_e, {\vec b}_\perp) =
\int \frac{d^2{\vec \Delta}_\perp}{(2\pi)^2}
e^{-i {\vec \Delta}_\perp \cdot {\vec b}_\perp }
H(x,0,-\vec{\Delta}_\perp^2).\label{eq:gpds_space} \ee
Here, the momentum transfer ${\vec \Delta}_\perp$  is the Fourier-conjugate to the impact parameter ${\vec b}_\perp$ and ${\vec b}_\perp$ corresponds to the displacement of the  electron $(e)$ from the transverse center of momentum  of the entire system $(e\bar{e})$. 

\section{Basis Light-Front Quantization}
The positronium bound-state problem is solved in a basis function approach on the light front \cite{Wiecki.2014}. In this approach, the longitudinal coordinate is confined in a
box of length $2\mathsf{L}$, $-\mathsf{L} \le x^- \le +{\mathsf L} $, with anti-periodic boundary condition for fermions. The longitudinal
momentum is discretized: $k^+ = (2j+1)\pi/\mathsf{L}, \; j=0,1,2,\cdots$. As the QED Hamiltonian is block diagonal for different $P^+$, we can fix it to be
$P^+ = 2K\pi/\mathsf{L}$, where $K$ is a positive integer. The longitudinal momentum fractions become, $x = (j+\frac{1}{2})/K, j=0,1,2,\cdots (K-1)$. It is clear that $K$ represents the resolution of the basis in the longitudinal direction. In the transverse direction, 2-dimensional (2D) harmonic oscillator (HO) functions are adopted as the basis. In terms of the dimensionless  transverse momentum variable $\vec{v}_\perp  (= \vec{k}_\perp / b \,\,\text{or}\,  \vec{\Delta}_\perp / b)$, the ortho-normalized 2D HO basis function reads
\begin{equation}
 \phi_{nm}(\vec v_\perp) = \sqrt{\frac{n!}{(n+|m|)!\pi}} e^{i m \theta} v^{|m|} e^{- v^2/2} L_n^{|m|}( v^2),\label{eq:HO_momentum}
\end{equation}
 where $v = |\vec v_\perp|$, $\theta = \arg \vec v_\perp$,
 $n$ and $m$ are the (2D) radial and angular quantum numbers,
 $L_n^\alpha(x)$ is the associated Laguerre polynomial, and $b$ is the HO basis scale with dimension of mass.

 For the spin degrees of freedom, two quantum numbers $\lambda_e$ and $\lambda_{\bar{e}}$ are used to label the helicities of the electron and positron, respectively. The momentum space LFWF used in Eq.~{\ref{eq:form_factor_intrinsic} reads
 \begin{equation}
\psi(\vec k_\perp, x, \lambda_e, \lambda_{\bar e}) =
\frac{1}{b \sqrt{x(1-x)}} \sum_{n, m} \langle n, m, x, \lambda_e, \lambda_{\bar e} | \psi\rangle
\phi_{nm} \left( \frac{\vec k_\perp}{b \sqrt{x(1-x)}} \right), \label{eq:LFWF_blfq}
\end{equation}
where $\langle n, m, x,\lambda_e,\lambda_{\bar e}|\psi\rangle$ is the LFWF in the BLFQ basis.

 The non-perturbative solutions for the LFWFs are provided by a recent BLFQ study \cite{Wiecki.2014}. Note that we have converted the non-perturbative solutions available in Ref.~\cite{Wiecki.2014} from single-particle coordinates to relative coordinates using the Talmi-Moshinsky (TM) transformation \cite{Talmi.1952}. Here, we exploit the fact that within the $N_{\text{max}}$ truncation (see below), the LFWFs preserve the factorization of the center of mass motion and the relative motion \cite{Maris:2013qma, Wiecki2013}. $n$ and $m$ in the LFWFs (Eq.~\ref{eq:LFWF_blfq}) are the quantum numbers in the relative coordinates.

 In order to make the numerical calculations feasible, the basis is made finite using truncation. In the relative coordinate for the $|e{\bar e}\rangle$ Fock-sector, the truncation on the transverse degree of freedom is applied as follows:
  \be 2n +|m|+1 \leq N_{\text{max}} .\ee
 In BLFQ, the total angular momentum $J$ is only an approximate quantum number, due to the breaking of the rotational symmetry by the Fock sector truncation and the basis
truncation. However, the total angular momentum projection for the system
\be  m_J = m + \lambda_e  + \lambda_{\bar{e}}\ee is conserved in our system.

Now, with the help of Eq.~\ref{eq:LFWF_blfq}, the GPDs (Eq.~\ref{eq:gpds_intrinsic}) in the BLFQ basis read

\be H(x, 0, -\vec\Delta_\perp^2)&=& \frac{1}{b^2\,x(1-x)}\,\,\,\,\,\,\,\,\, \sum_{\mathclap{n,n',m,\lambda_e, \lambda_{\bar e}}} \langle \psi | n', m, x, \lambda_e,\lambda_{\bar e} \rangle \langle n, m, x, \lambda_e,\lambda_{\bar e} | \psi \rangle \\ \nonumber
&&\times \int d^2 \vec k_\perp  \phi^*_{n' m}\biggr(\frac{{\vec k'}_\perp}{b\sqrt{x(1-x)}}\biggr)  \phi_{n m}\biggr(\frac{{\vec k}_\perp}{b\sqrt{x(1-x)}}\biggr)\\ \nonumber
&=& \frac{1}{b^2\,x(1-x)}\,\,\,\,\,\,\,\,\, \sum_{\mathclap{n,n',m,\lambda_e, \lambda_{\bar e}}} \langle \psi | n', m, x, \lambda_e,\lambda_{\bar e} \rangle \langle n, m, x, \lambda_e,\lambda_{\bar e} | \psi \rangle \\
&&\times \int d^2 \vec k_\perp  \phi^*_{n' m}\biggr(\frac{{\vec k}_\perp + \frac{(1-x){\vec \Delta}_\perp}{2 }} {b\, \sqrt{x(1-x)}} \biggr) \phi_{n m}\biggr(\frac{{\vec k}_\perp - \frac{(1-x){\vec \Delta}_\perp}{2 }} {b\, \sqrt{x(1-x)}} \biggr)
, \ee
where $b$ is the HO basis scale with dimension of mass}.
Note that in the last step, we have applied a shift in integration variables. Now, the  integral over the product
of the two HO functions with different arguments can be simplified by using the
TM coefficients for the 2D-HO functions [36] to reduce it to an integral
over one HO function.  Thus, one can write

\be H(x, 0, -\vec\Delta_\perp^2)
&=& \sqrt{\pi} \sum_{\mathclap{n,n',m,\lambda_e, \lambda_{\bar e}}}  (-1)^{n+n'+|m|}
\langle \psi | n', m, x, \lambda_e,\lambda_{\bar e} \rangle
\langle n, m, x, \lambda_e,\lambda_{\bar e} | \psi \rangle \label{eq:gpds_blfq}\\ \nonumber
&& \times \sum_{\nu} \mathcal M^{N,0,\nu,0}_{n,m,n',-m} (-1)^\nu \phi_{\nu 0}\biggr(\sqrt{\frac{1-x}{2x}}\frac{\vec \Delta_\perp}{b}\biggr),
\ee
where $\mathcal M^{N,0,\nu,0}_{n,m,n',-m}$ are TM coefficients used to separate the center of mass part and the relative part in the basis functions \cite{Talmi.1952}, $N = n+n'-\nu+|m|$, $0 \le \nu \le n+n'+|m|$, $\phi_{nm}({\vec v}_\perp)$ is the 2D-HO basis function in momentum space, see Eq.~\ref{eq:HO_momentum}, and $x = (j+\frac{1}{2})/K, j=0,1,2,\cdots (K-1)$. Readers are referred to Ref.~\cite{Wiecki.2014} for details  of the BLFQ approach and center of mass factorization.

Integrating Eq.~\ref{eq:gpds_blfq} with respect to $x$, one can get the FFs in BLFQ basis, i.e.
\be
F(t) = \int_0^1 dx \, H(x, 0, -\vec\Delta_\perp^2) \approx \sum_{j=0}^{K-1}\,\frac{1}{K}\, H\biggr(\frac{2j+1}{2K}, 0, -\vec\Delta_\perp^2\biggr), \label{eq:form_factor_blfq}\ee
where the approximation becomes exact in the continuum limit $K \rightarrow \infty$.

Inserting Eq.~\ref{eq:gpds_blfq} in Eq.~\ref{eq:gpds_space}, $q(x, {\vec b_\perp})$ in the BLFQ reads
\be  q(x,{\vec b}_\perp)&=&\frac{b^2}{\sqrt{\pi}}\frac{x}{(1-x)}\,\,\;\;\;\sum_{\mathclap{ n,n',m,\lambda_e, \lambda_{\bar e}}}  (-1)^{n+n'+|m|}
\langle \psi | n', m, x, \lambda_e,\lambda_{\bar e} \rangle
\langle n, m, x, \lambda_e,\lambda_{\bar e} | \psi \rangle \nonumber \\
&& \times \sum_{\nu} \mathcal M^{N,0,\nu,0}_{n,m,n',-m} (-1)^\nu \widetilde{\phi}_{\nu 0}\biggr(\sqrt{\frac{2x}{1-x}} {\,b \, \vec b}_\perp\biggr),\label{eq:gpds_space_blfq} \ee
where $\widetilde\phi_{nm}(b\,{\vec b_\perp})$ is the Fourier transform of Eq.~\ref{eq:HO_momentum}.

\begin{figure}
\begin{tabular}{cc}
\subfloat[$1^1S_0 (0^{-+})$ with $b=0.5 m_e$]{\includegraphics[scale=0.39]{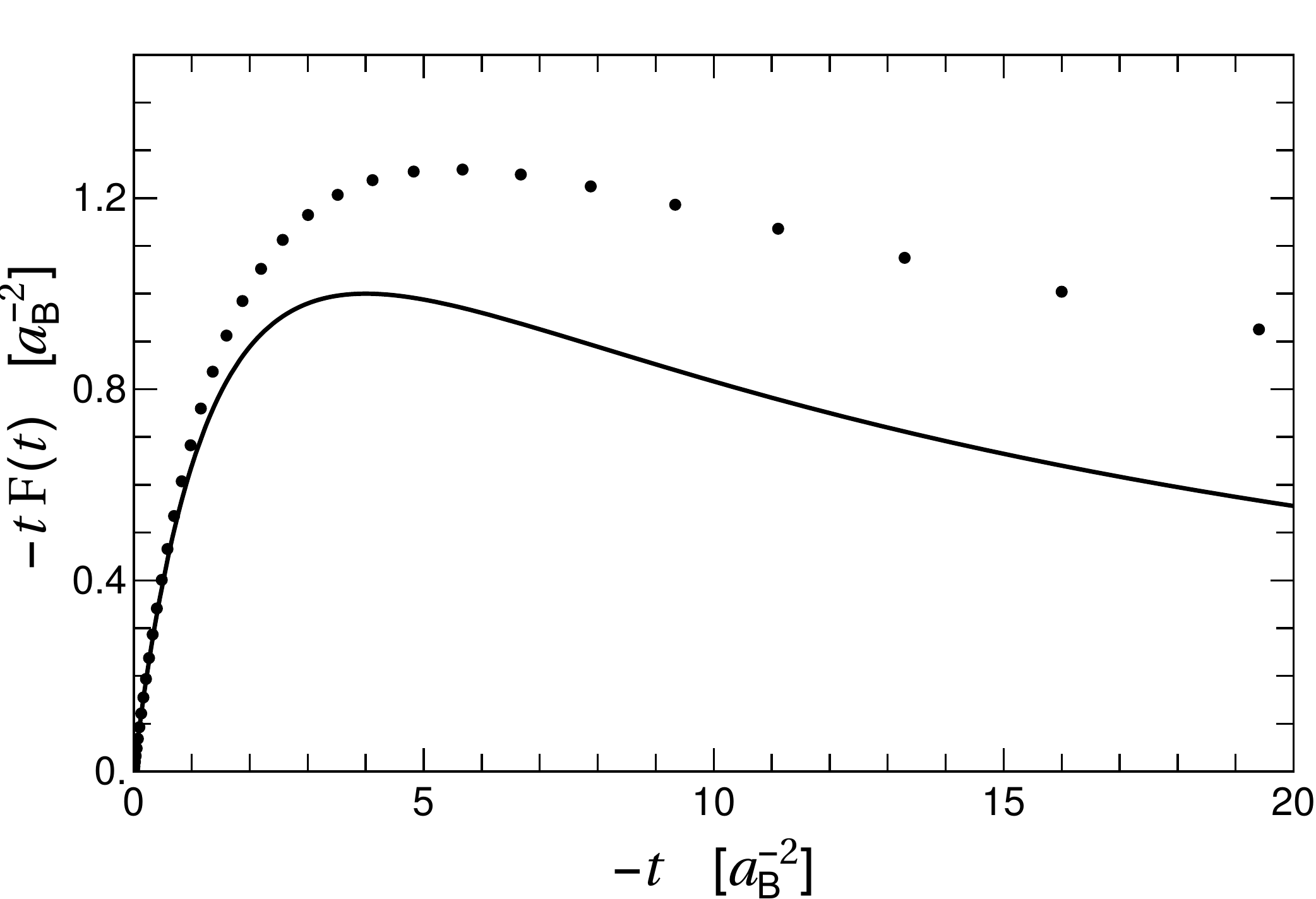}}\\
 \subfloat[$1^3S_1 (1^{--})$ with $b=0.5 m_e$]{\includegraphics[scale=0.39]{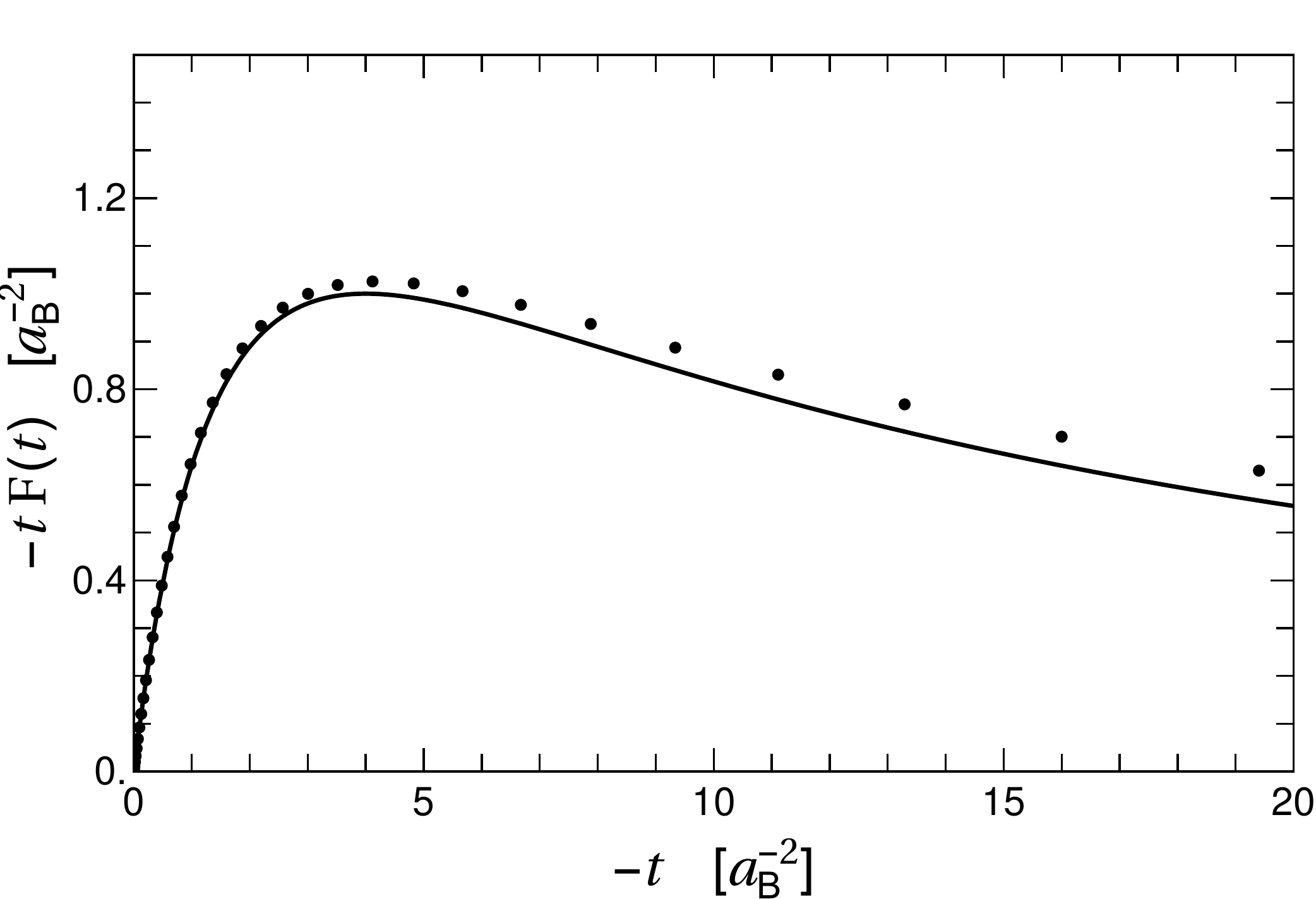}} \\
 \end{tabular}
\begin{tabular}{cc}
\subfloat[$2^1S_0 (0^{-+})$ with $b=0.1 m_e$]{\includegraphics[scale=0.39]{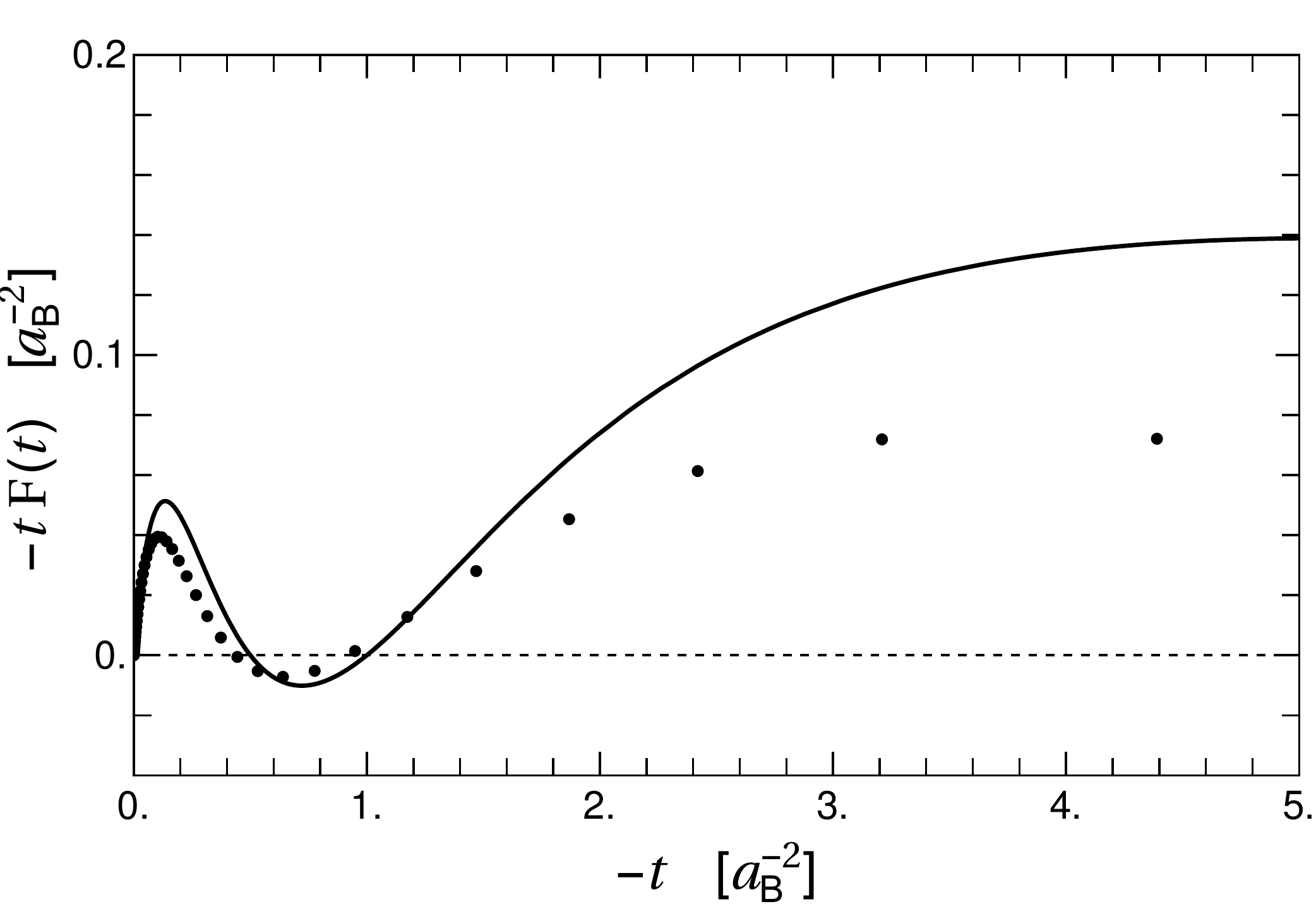}}\\
\subfloat[$2^3P_0 (0^{++})$ with $b=0.1 m_e$]{\includegraphics[scale=0.39]{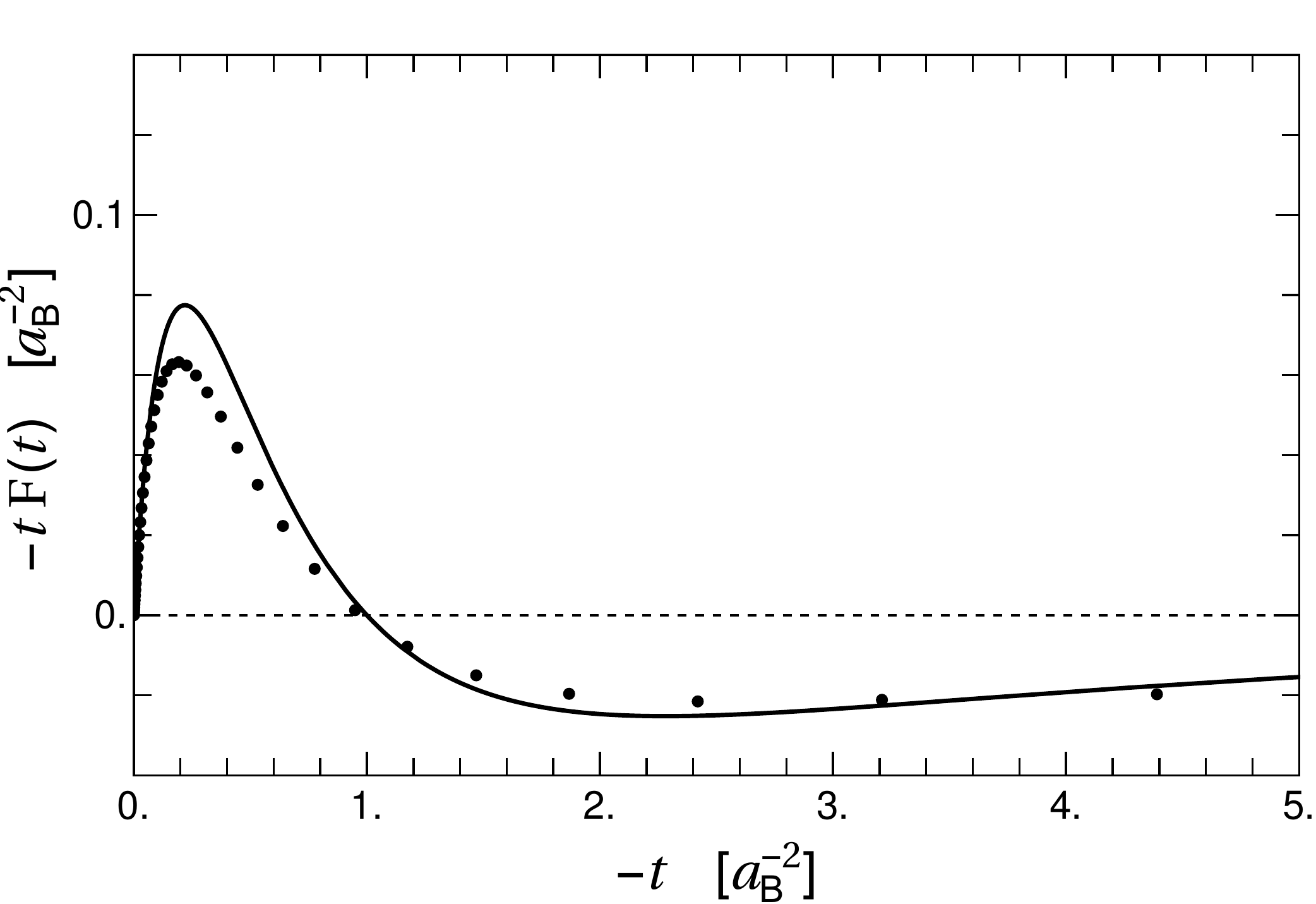}} \\
\end{tabular}
\caption{ $-t F(t)$ vs $-t$ for the four low-lying  bound states of positronium with $N_{\text{max}}=31$, $K=61$, $m_J =m_J' = 0$, coupling constant $\alpha=0.3$, and photon mass $\mu = 0.02m_e$. The ``dotted line" represents the positronium FF calculations $F(t)$ in BLFQ basis (Eq.~\ref{eq:form_factor_blfq}) and ``solid line" represents the FF calculations $F^{NR}_{\ell=0}(q^2)$ from non-relativistic quantum mechanics  (Eq.~\ref{eq:NRQM}). Note $t = -\Delta_\perp^2$ and $a_\textsc{b}$ = $1/( \alpha m_e)$ is the Bohr radius and $b$ is the basis scale for the HO functions.\label{fig:Q_form_factor}}
\end{figure}

\begin{figure}
\includegraphics[scale=0.52]{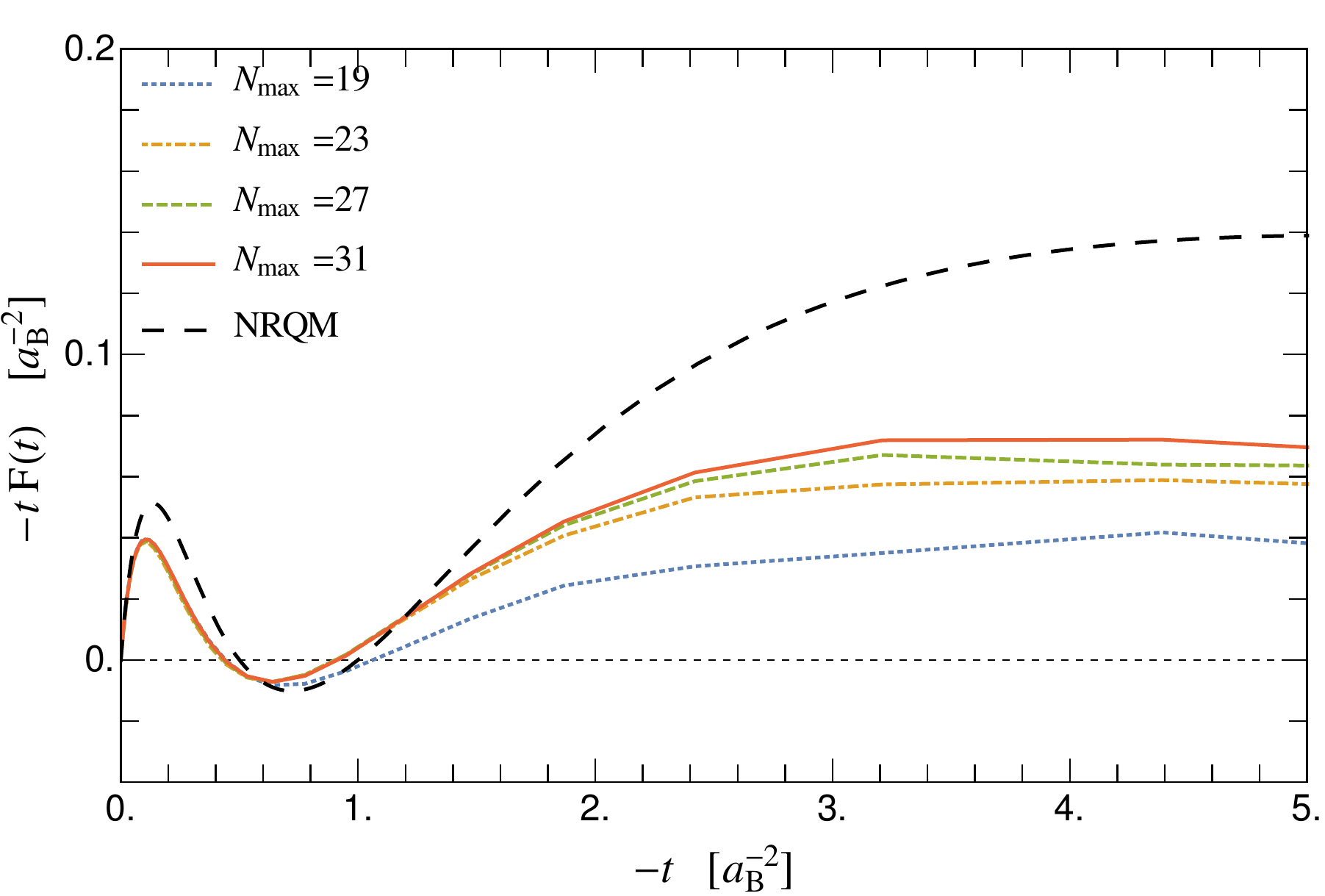}
\caption{ $-t F(t)$ vs $-t$ for $2^1S_0 (0^{-+})$ with different $N_{\text{max}}$. The results are calculated at $K=61$, $m_J =m_J' = 0$, coupling constant $\alpha=0.3$, $b=0.1 m_e$, and photon mass $\mu = 0.02m_e$. Note $t = -\Delta_\perp^2$, $a_\textsc{b}$ = $1/( \alpha m_e)$ is the Bohr radius, and $b$ is the basis scale for the HO functions.\label{fig:convergence}}
\end{figure}

\section{Results and discussion}
We now present and discuss our results for FFs and GPDs obtained in the BLFQ approach beginning with the FFs  for the four low-lying bound states of positronium in Fig.~\ref{fig:Q_form_factor}. For FFs, the results are calculated with fixed photon mass $\mu = 0.02 m_e$, where $m_e$ is the mass of the electron, and with $N_{\text{max}} = 31, K = 61$. The small photon mass $\mu$ was introduced as a regulator in the two-body effective interaction in Ref.~\cite{Wiecki.2014}. The basis scale $b$ is chosen to minimize the ground-state energy at the given $N_{\text{max}}$ and $K$ truncation for the given regulator $\mu$ and the given coupling constant $\alpha =0.3$. We present result in units of the Bohr radius $a_\textsc{b}$ = $1/( \alpha m_e)$.
We compare our positronium FFs calculated in BLFQ with the Non-Relativistic Quantum Mechanics (NRQM) FFs based on the multipole expansion of the one-body charge density. With suitable changes in the NRQM, we can adapt the one-body charge density calculated from the wave functions available for the hydrogen atom.

In NRQM, one can define non-relativistic FFs for different states by \cite{Book:NRQM}
\begin{equation}
\begin{split}
F^{NR}_\ell(q^2) \equiv& \sqrt{4\pi} i^\ell  \langle n, L, m_L  |j_\ell(\half q r) Y_\ell^0(\hat r) | n,
L, m_L \rangle \\
=& \sqrt{4\pi} i^\ell \int d^3r \, j_\ell(\half qr) \rho_L(\vec r) Y_\ell^0(\hat r) \quad (\ell = 0, 1, \cdots, 2L),\label{eq:NRQM}
\end{split}
\end{equation}
where $F^{NR}_\ell(q^2)$ are the multipole FFs,  ${\vec q}$ is the momentum transfer, $n$ is the principal quantum number, $L$  is the total orbital angular
momentum $({\vec L} = {\vec J}-{\vec S})$ and $m_L$ is its magnetic projection, $J$ is the total angular momentum, $\rho_L({\vec r}) \triangleq \psi^*({\vec
r})\psi({\vec r})=\langle n, L, m_L|\vec r\rangle\langle \vec r|n, L, m_L\rangle$ is the coordinate space one-body charge density,
$Y_\ell^{m}(\hat r)$ is the spherical harmonics, and $j_\ell(z)$ is the spherical Bessel function of the first kind. In our present work, the FFs $F^{NR}_\ell(q^2)$ with $\ell=0$ are compared with the FFs $F(t)$  in BLFQ basis (Eq.~\ref{eq:form_factor_blfq}).

In Fig.~\ref{fig:Q_form_factor}, one of the features that both FF calculations  have in common is the formation of the nodes in $N=2$ states of positronium at lower $|t| = |\Delta^2|$.
Furthermore, one can easily recognize the FF calculations in BLFQ (``dotted line") are consistent with the FF calculations from the NRQM  (``solid line") at small momentum transfer. The differences between the BLFQ and NRQM results for the FFs in Fig.~\ref{fig:Q_form_factor} begin to be evident around $|t|\sim 2 a_\textsc{b}^{-2}$ for the nodeless states and at $|t| \sim 0.2 a_\textsc{b}^{-2}$ for the states with radial nodes. As we mentioned before for $J=1$, the FF we have calculated is $I_{0,0}$.

Here, we also investigate the convergence of the FF with respect to $N_{\text{max}}$ motivated by the observation that the calculations in Ref.~\cite{Wiecki.2014} showed that the mass spectrum is more sensitive  to $N_{\text{max}}$ as compared to $K$ or $\mu$. Therefore, our FF and GPD results are calculated with fixed $K$ and $\mu$. Note that $N_{\text{max}}$ is the parameter governing the truncation in the transverse direction and we might surmise that the FF defined through momentum transfer in the transverse plane is therefore more sensitive to this truncation.  Fig.~\ref{fig:convergence} shows the convergence of the FF calculations with respect to $N_{\text{max}}$ for $2^1S_0 (0^{-+})$  keeping other regulators, $K=61$, and $\mu$ = $0.02 m_e$ fixed. As may be expected, Fig.~\ref{fig:convergence} suggests that the FF calculations have better  $N_{\text{max}}$ convergence at lower $|t|=|{\Delta}^2|$ since the higher momentum transfers probe details of the charge density requiring higher HO basis states for accurate descriptions.
\begin{figure}
\begin{tabular}{cc}
\subfloat[$1^1S_0 (0^{-+})$ with $b=0.5 m_e$]{\includegraphics[scale=0.37]{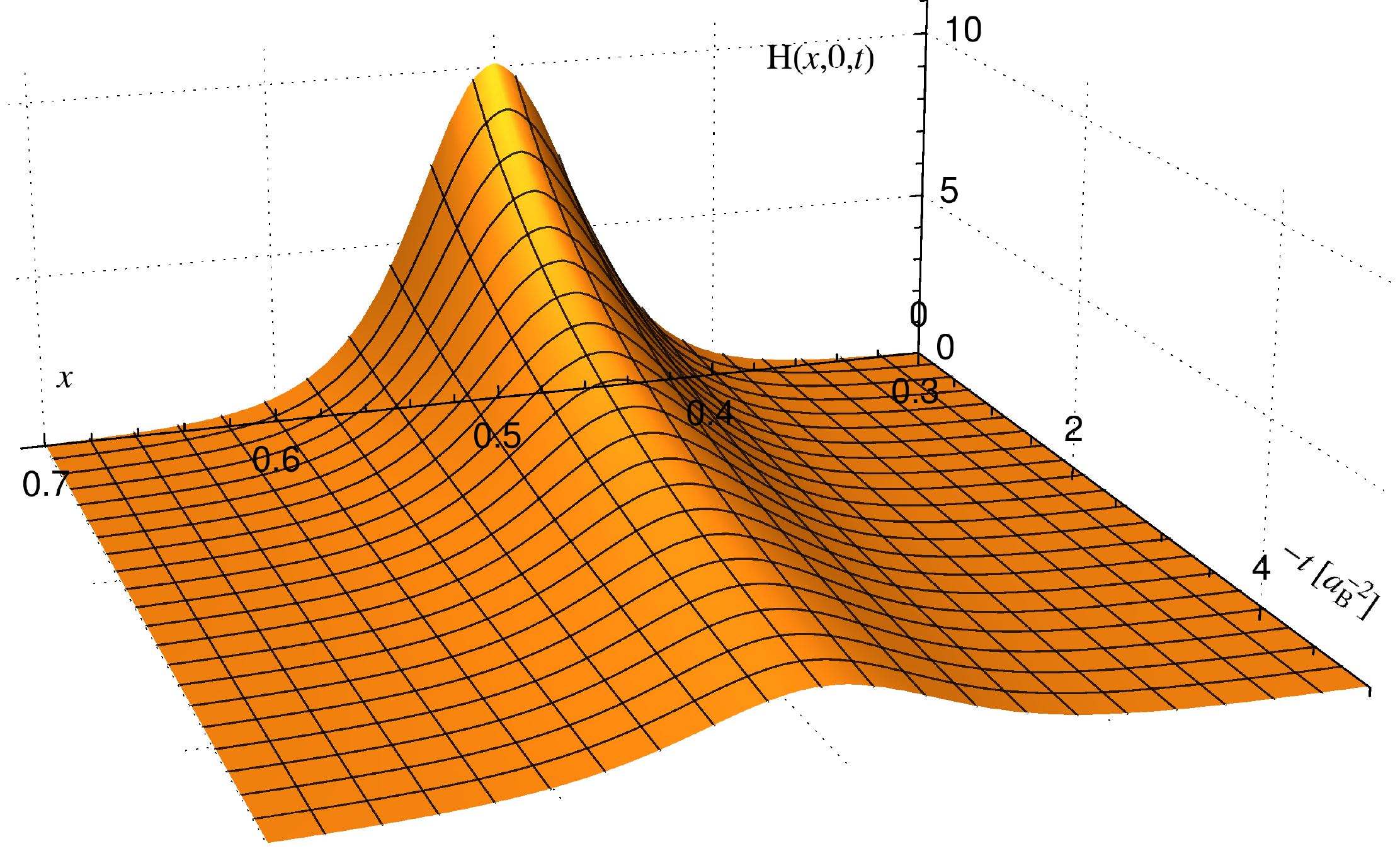}}\\
\subfloat[$1^3S_1 (1^{--})$ with $b=0.5 m_e$]{\includegraphics[scale=0.37]{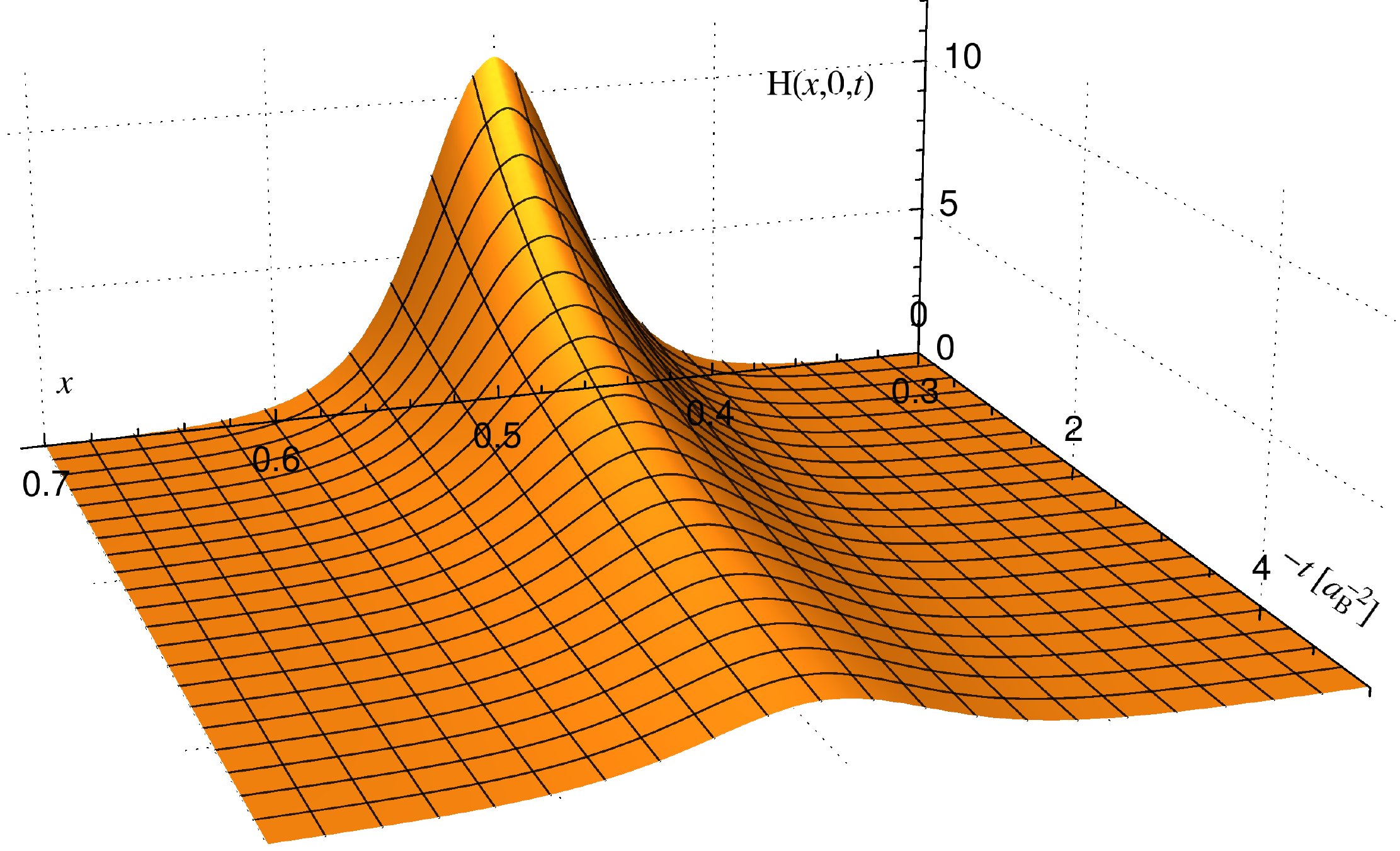}} \\
\end{tabular}
\begin{tabular}{cc}
\subfloat[$2^1S_0 (0^{-+})$ with $b=0.1 m_e$]{\includegraphics[scale=0.37]{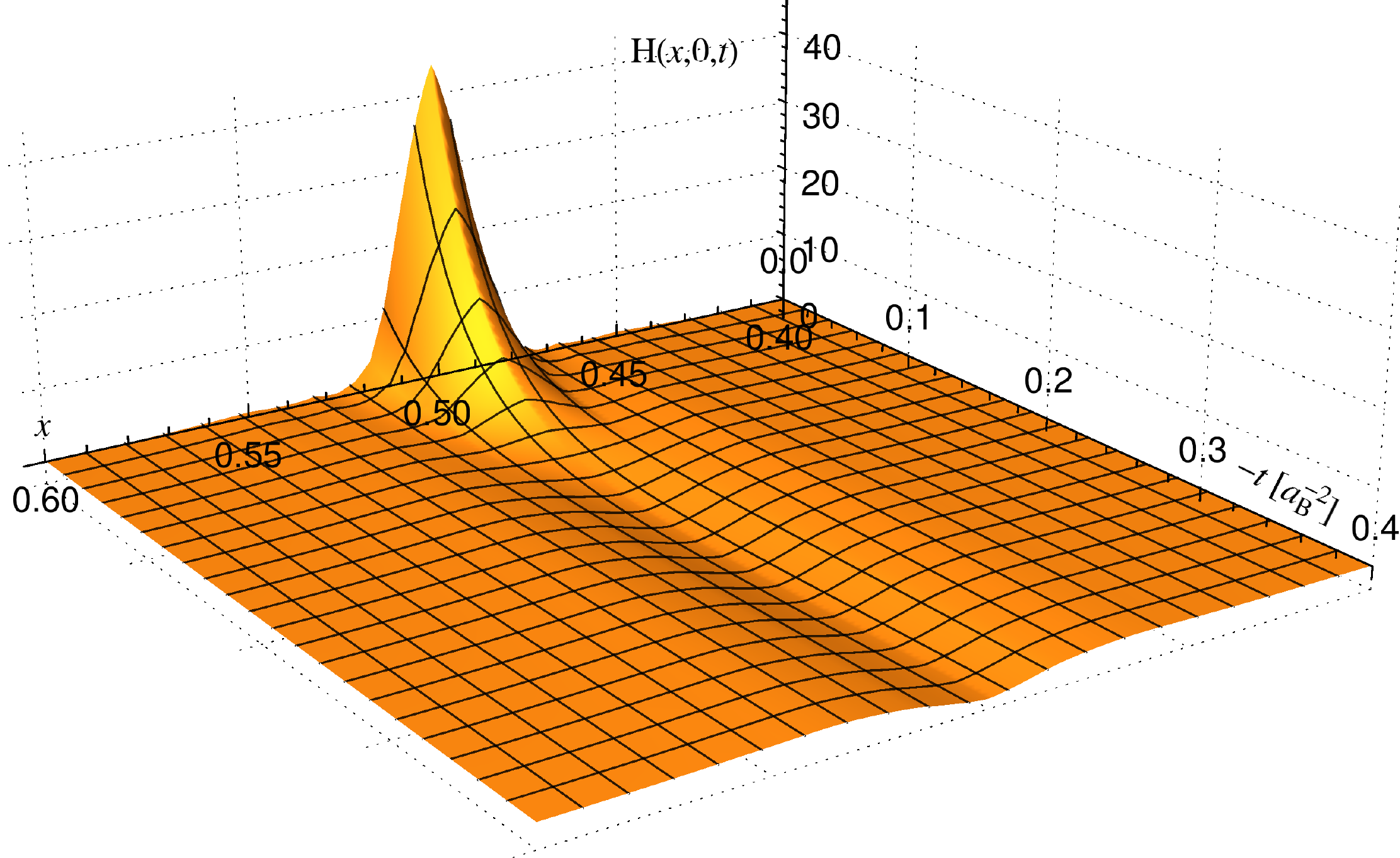}}\\
\subfloat[$2^3P_0 (0^{++})$ with $b=0.1 m_e$]{\includegraphics[scale=0.37]{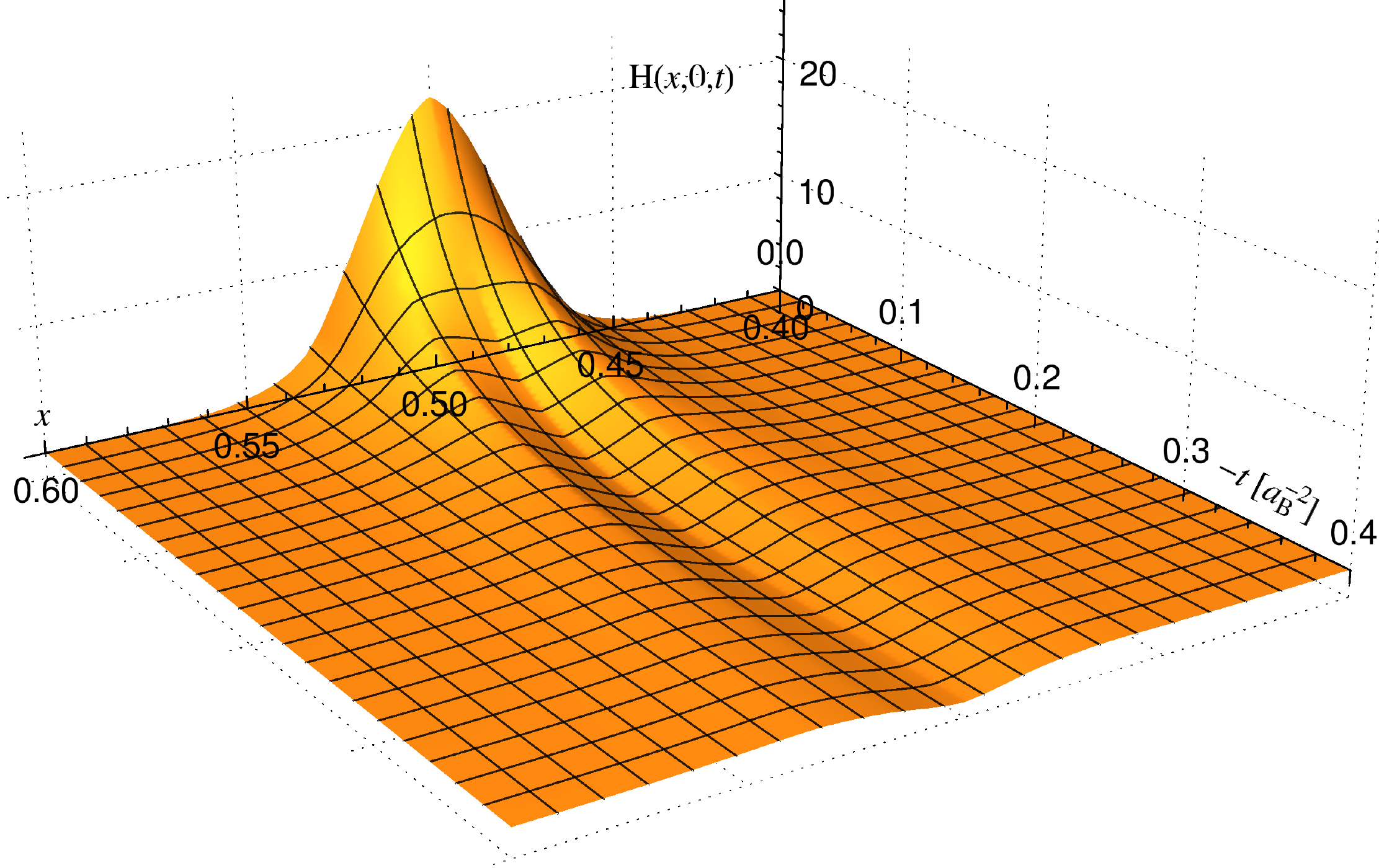}} \\
\end{tabular}
\caption{3D plot of helicity non-flip GPDs  $H(x, \xi=0, t=-\Delta_\perp^2)$ (Eq.~\ref{eq:gpds_blfq}) for the four low-lying bound states of positronium with $N_{\text{max}}=31$, $K=61$, $m_J =m_J' = 0$, coupling constant $\alpha=0.3$,  and photon mass $\mu = 0.02m_e$. Note $a_\textsc{b}$ = $1/( \alpha m_e)$ is the Bohr radius and $b$ is the basis scale. \label{fig:gpds}}
\end{figure}

Next, we present GPDs for the four low-lying bound states of positronium in Fig.~\ref{fig:gpds} for fixed $N_{\text{max}}= 31,\, K = 61,\,$ and $\mu = 0.02 m_e$. The  parton distribution in each case shown in Fig.~\ref{fig:gpds} is  peaked at $x=0.5$ which reflects the symmetry between the electron and the positron in the positronium system. The $t$-dependence of the positronium  GPDs provides insights into the non-perturbative structure of the system. It is interesting to note that the $x$-dependence changes character from low $|t|$ to high $|t|$ for some states as seen in Fig.~\ref{fig:gpds}(c) and Fig.~\ref{fig:gpds}(d). Furthermore, in momentum space, the decaying trend of the GPDs is more rapid with increasing $|t|$  for $N=2$ compared to $N=1$. This signifies the radial extension of the former states is broader than that of the latter states (see the impact-parameter dependent GPDs below). This fact is consistent with NRQM because positronium in a radially excited state is more loosely bound with a longer coordinate space tail to its wave function.

The impact-parameter dependent GPDs $q(x,{\vec b}_\perp)$ for our four selected positronium states are  presented in Fig.~{\ref{fig:q_bperp} for fixed $N_{\text{max}}= 31,\, K = 61,\,$ and $\mu = 0.02 m_e$. The plots show the distribution of an electron carrying momentum fraction $x$ in the transverse plane as a function of ${\vec b}_\perp$. As we know ${\vec b}_\perp = (1-x) {\vec r}_\perp$, where ${\vec r}_\perp$ is the transverse separation between the electron and the positron \cite{mb:IJMP,mb:GPD, Diehl:2003ny, Soper:1976jc},  the distribution is generally asymmetric with respect to $x=0.5$ for ${\vec b}_\perp \ne0$. It is interesting to note the more complex landscapes appearing in Fig.~\ref{fig:q_bperp}(c) and Fig.~\ref{fig:q_bperp}(d) where bi-modal distributions in $x$ at low ${\vec b_\perp}$ transform into single mode distributions in $x$ at moderate ${\vec b_\perp}$.
\begin{figure}
\begin{tabular}{cc}
\subfloat[$1^1S_0 (0^{-+})$ with $b=0.5 m_e$]{\includegraphics[scale=0.36]{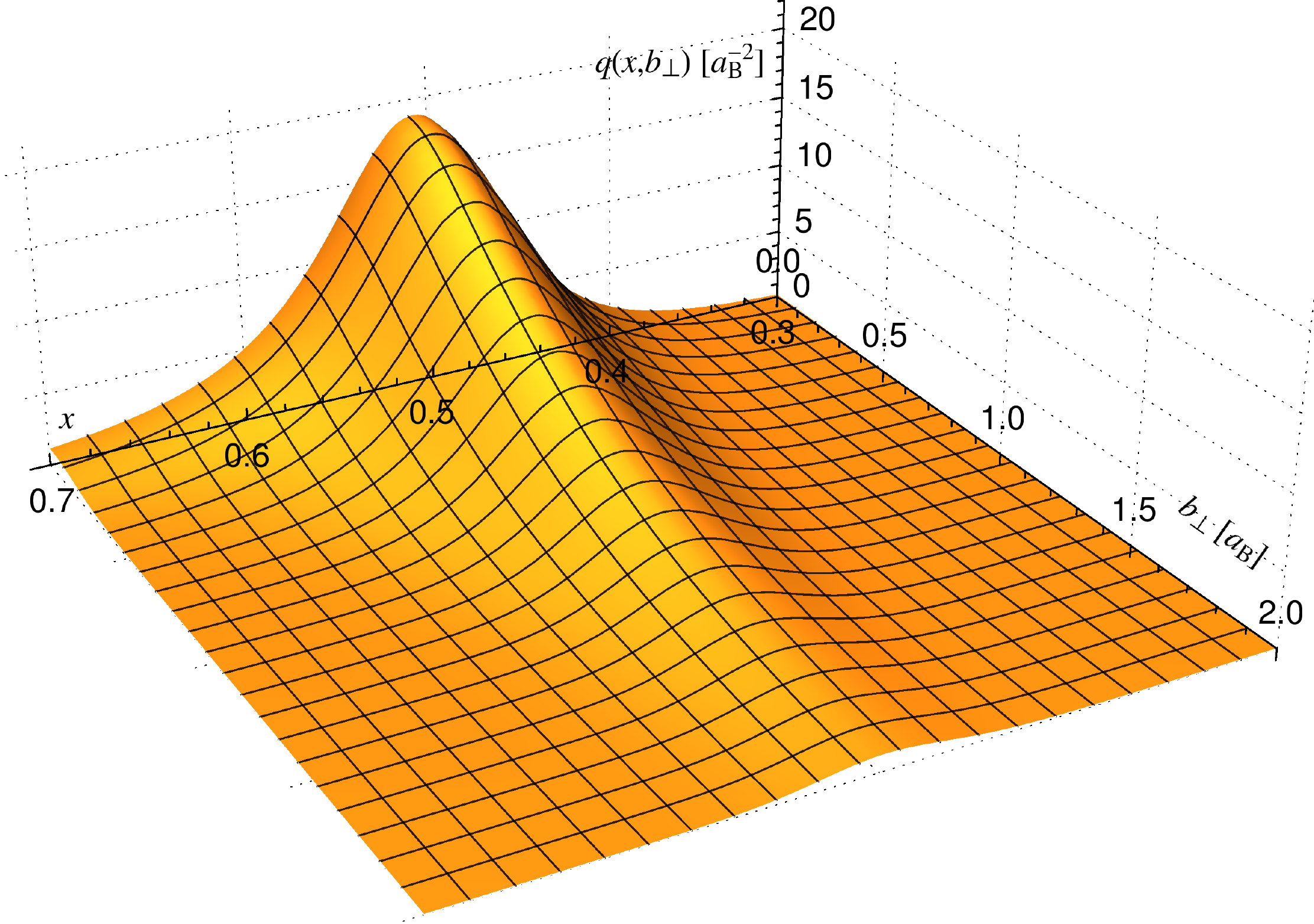}}\\
 \subfloat[$1^3S_1 (1^{--})$ with $b=0.5 m_e$]{\includegraphics[scale=0.36]{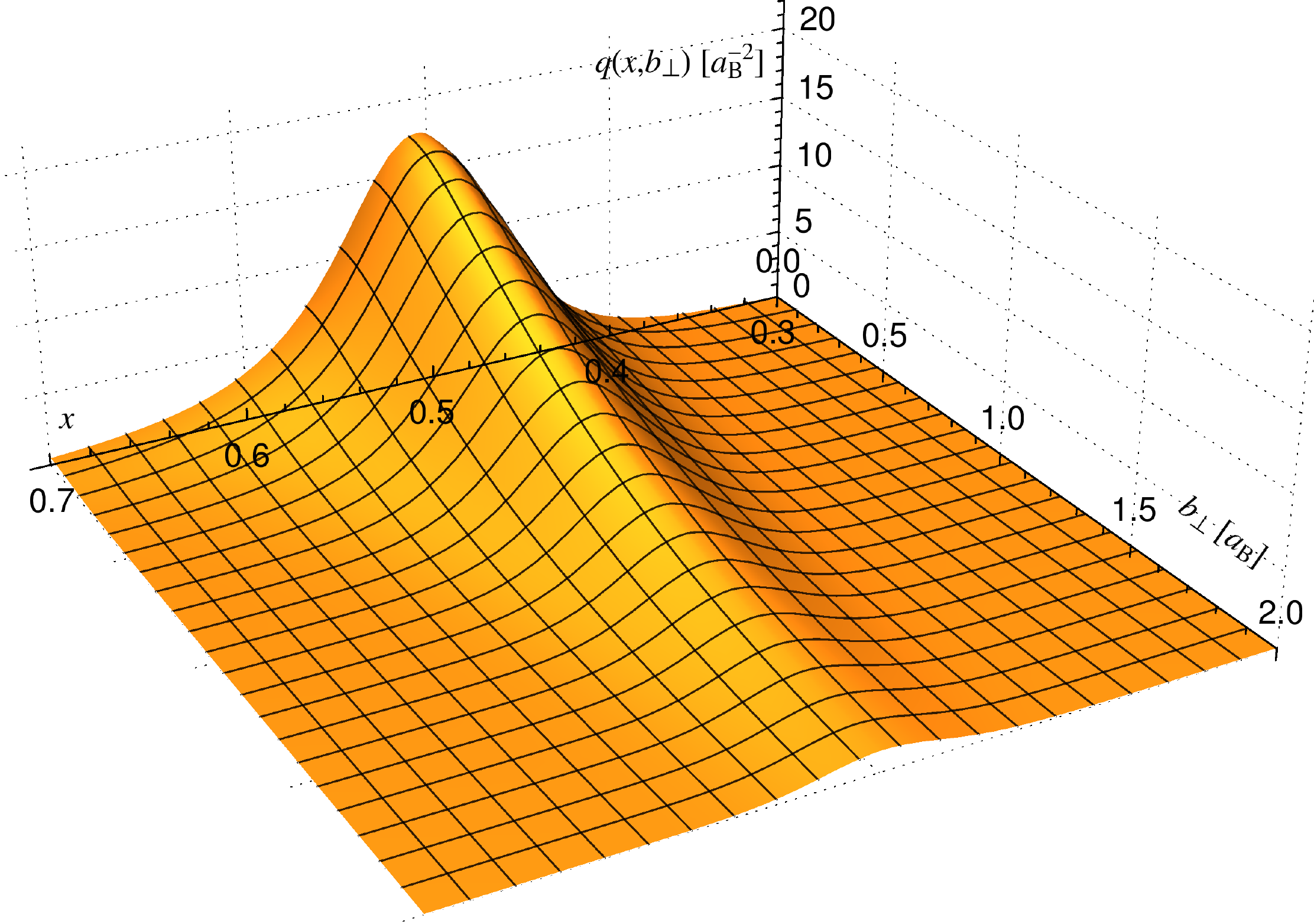}} \\
 \end{tabular}
\begin{tabular}{cc}
\subfloat[$2^1S_0 (0^{-+})$ with $b=0.1 m_e$]{\includegraphics[scale=0.39]{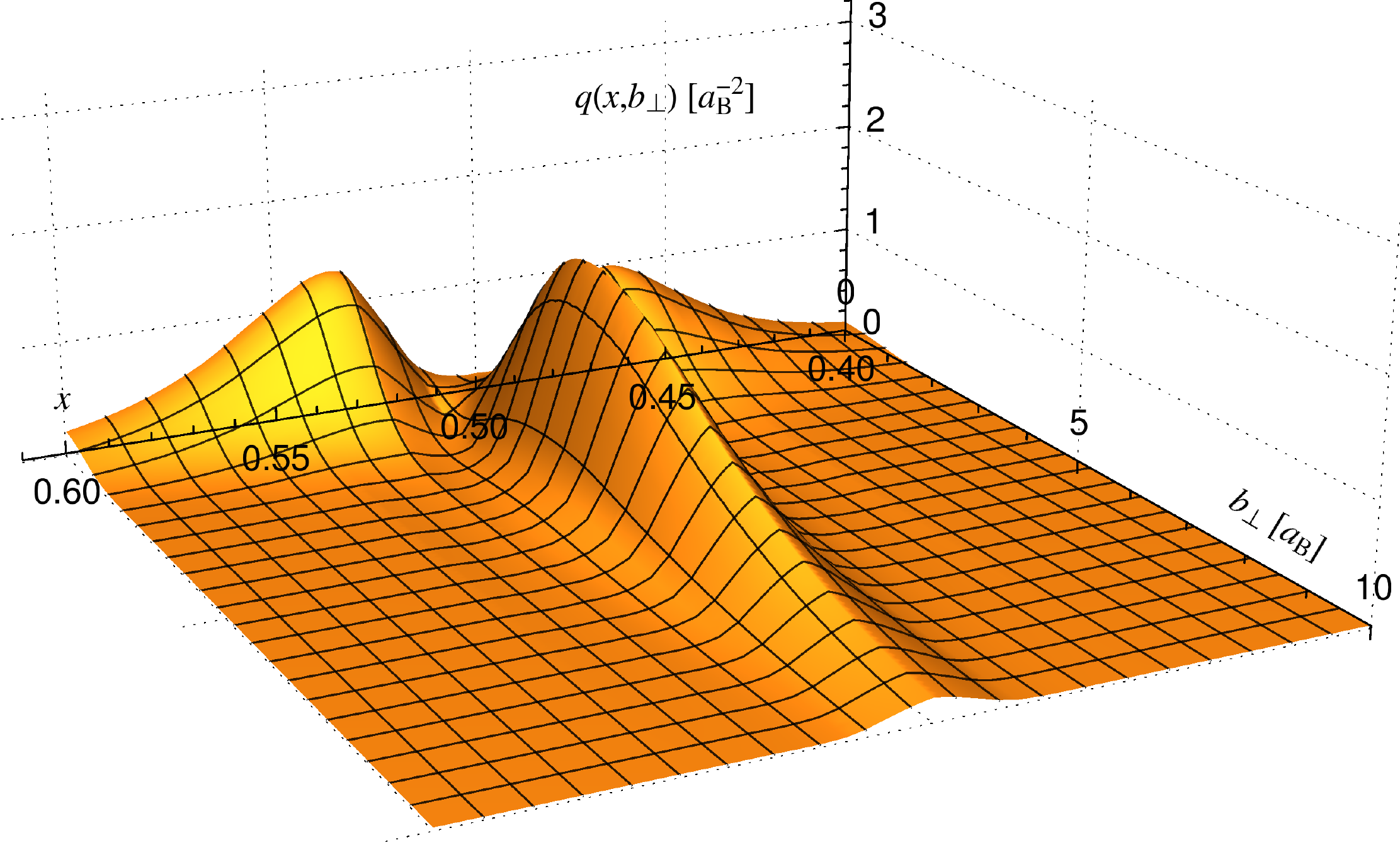}}\\
\subfloat[$2^3P_0 (0^{++})$ with $b=0.1 m_e$]{\includegraphics[scale=0.39]{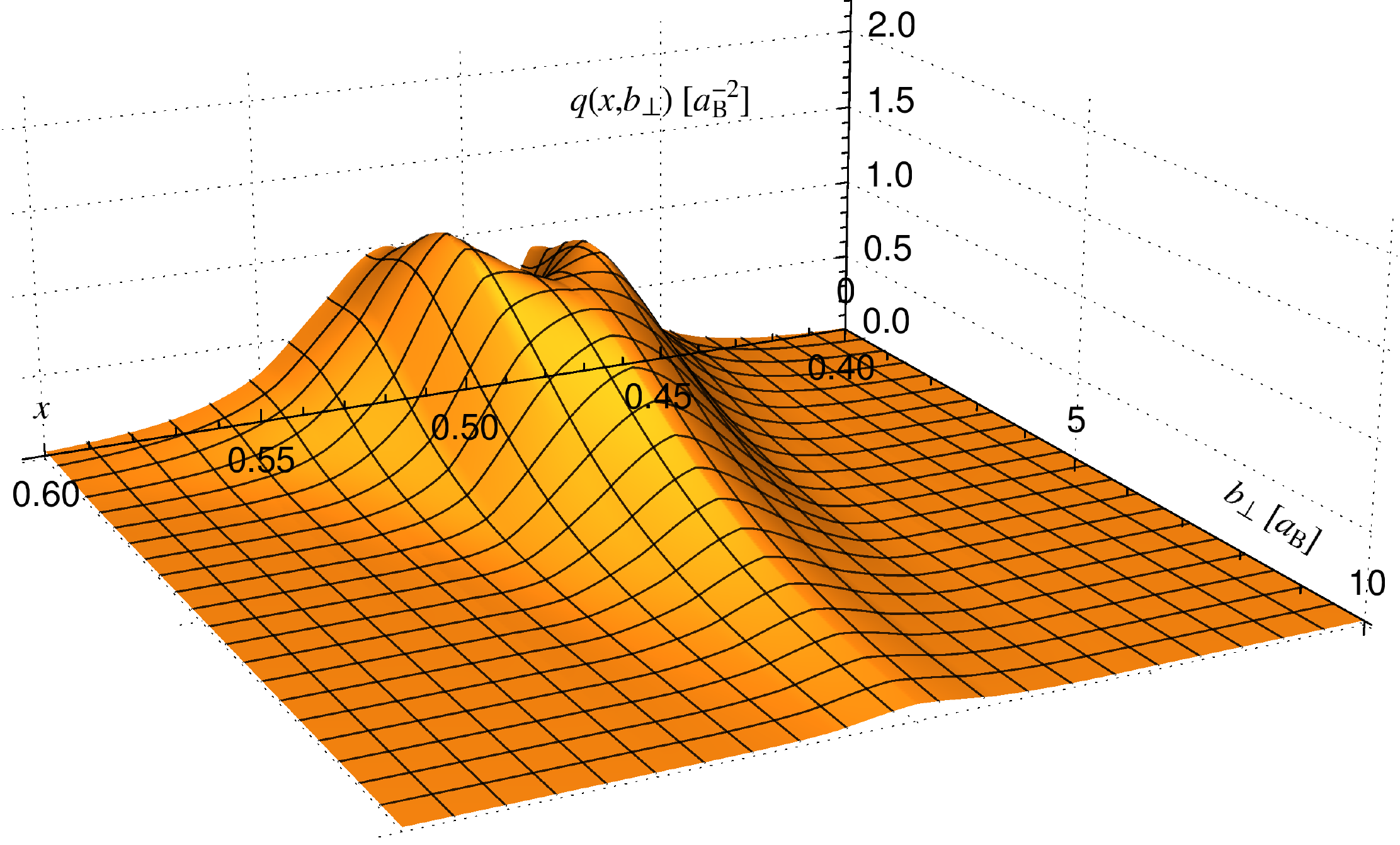}} \\
\end{tabular}
\caption{Impact-parameter dependent GPDs $q(x,{\vec b}_\perp)$ (Eq.~\ref{eq:gpds_space_blfq}) for the four low-lying  bound states of positronium with $N_{\text{max}}=31$, $K=61$, $m_J =m_J' = 0$, coupling constant $\alpha=0.3$,  and photon mass $\mu = 0.02m_e$. Note $a_\textsc{b}$ = $1/( \alpha m_e)$ is the Bohr radius and $b$ is the basis scale.  \label{fig:q_bperp}}
\end{figure}

\section{Summary and Outlook}
We have calculated the GPDs, FFs, and impact-parameter dependent GPDs for the model fermion-antifermion problem of positronium at strong coupling in the BLFQ approach. We have compared FFs calculated in BLFQ with those from the one-body density in momentum space in NRQM. They agree reasonably well in low-momentum transfer region, as may be expected. We have also studied the convergence with respect to $N_{\text{max}}$ for selective observables. Convergence is reasonably well-established in the low-momentum transfer region.

 We can extend the present work to higher spin states ($J\ge 1$) and compute different physical FFs such as magnetic FFs and quadrupole FFs. Such results can further be compared with those of quarkonium systems such as $c\bar{c}$ and $b\bar{b}$ that have been solved in the BLFQ framework \cite{Li:2015zda}. Moreover, visualizing the simpler positronium system at strong coupling in 3 dimensions provides us with benchmark cases that will help us better understand features that may arise with  the hadronic structure of mesons in the non-perturbative framework. Our  ultimate goal is to apply this non-perturbative method to QCD to compute and study observables such as FFs and GPDs of the proton.

\section{Acknowledgements}
It is a pleasure to thank P. Wiecki, V.A. Karmanov, and G. Chen for helpful discussions. This work was supported in part by the Department of Energy under Grant Nos. DE-FG02-87ER40371 and DESC0008485 (SciDAC-3/NUCLEI). X. Zhao is supported by the new faculty startup funding by the Institute of Modern Physics, Chinese Academy of Sciences. Computational resources were provided by the National Energy Research Scientific Computing Center (NERSC), which is supported by the Office of Science of the U.S. Department of Energy under Contract No. DE-AC02- 05CH11231.

\end{document}